# A Time Decomposition and Coordination Strategy for Power System Multi-Interval Operation

Farnaz Safdarian *Student Member, IEEE,* Okan Ciftci, *Student Member, IEEE,* and Amin Kargarian, *Member, IEEE*

*Abstract*— This paper presents a time decomposition strategy to reduce the computational complexity of power system multi-interval operation problems. We focus on the economic dispatch problem. The considered scheduling horizon is decomposed into multiple smaller sub-horizons. The first time interval of each sub-horizon is modeled as the coupling interval between two consecutive sub-horizons. The interdependencies between the sub-horizons are mathematically modeled using ramp rates of generating units. A distributed coordination strategy, which is based on auxiliary problem principle, is developed to coordinate the economic dispatch solutions of the sub-horizons to find an optimal solution for the whole operation horizon. We also propose an initializing technique to start the iterative coordination algorithm from a good-enough point. This technique enhances the convergence rate significantly. The proposed algorithm is deployed to solve a week-ahead economic dispatch problem on the IEEE 118-bus system, and promising results are obtained.

## I. INTRODUCTION

MANY power systems analysis and decision-making problems are based on formulation and solution of large-scale multi-interval optimization models. One of these decision-making problems, which is studied in this paper, is a multi-interval economic dispatch (ED)[1, 2]. Depending on the application and type of analysis, ED's time horizon could be one day, one week, etc. [3]. In such an optimization problem, the size of the search space is large, and as a result, the computational burden will be high. Another challenge is intertemporal constraints that interconnect decisions made in a time interval to decisions made in other intervals. Limits of ramping capabilities of generating units [4] are intertemporal constraints of the multi-interval ED problem. These constraints increase the complexity of ED.

Various techniques have been presented in the literature to reduce the computational complexity and costs of the ED problem. One of the most popular techniques is to deploy decomposition algorithms along with distributed optimization methods [5-7]. Most of these techniques work based on the geographical decomposition to create several smaller subsystems than the original system [6, 8-14]. Since the subsystems are coupled, for instance, through voltages of buses at the boundaries of the subsystems, distributed/decentralized algorithm are applied to coordinate solutions of the subproblems [12]. We call such forms of geographical decompositions as vertical decompositions.

Although decomposing the system geographically potentially reduces the size and computational burden of an optimization problem, it does not deal with the intertemporal constraints, which complicate the decision-making process in the ED problem. A decomposition-coordination strategy that reduces the size of the optimization and mitigates the impact of intertemporal constraints is, potentially, a promising approach to solve the multi-interval ED problem. Intuitively, such a decomposition can be implemented over the considered operation horizon. We call this strategy as a horizontal decomposition since it decomposes the optimization with respect to the intertemporal constraints that interconnect the optimization time intervals horizontally.

In this paper, we aim to solve an economic dispatching (ED) problem, which is multi-interval decision-making model, by decomposing the optimization problem horizontally. We propose a time decomposition strategy to divide ED into several smaller optimization subproblems than the original ED problem. Each subproblem is formulated to solve a sub-horizon of the whole considered operation horizon. The consecutive sub-horizons are coupled via a set of complicating intertemporal constraints. We model the first time interval of each sub-horizon as the coupling intervals between that sub-horizon and the previous sub-horizon. This coupling interval models the ramping limits of the generating units for transition from sub-horizon $T_i$ to sub-horizon $T_{i+1}$. To coordinate the solutions of the sub-horizon and ensure the feasibility of the results from the perspective of the system components, we introduce a distributed coordination strategy that is based on auxiliary problem principle. In addition, we propose an initialization technique to find a set of good initial values and speed up the convergence rate of the distributed algorithm. The proposed time decomposition and coordination strategy is

This project was supported by the Louisiana Board of Regents under grant LEQSF (2016-19)-RD-A-10.
The authors are with the department of Electrical and Computer Engineering, Louisiana State University, Baton Rouge, LA 70803 USA (email: fsafda1@lsu.edu, ociftc1@lsu.edu, kargarian@lsu.edu).

applied to solve a week-ahead ED problem on the IEEE 118-bus system. The algorithm provides promising results with the aim of computational time reduction.

The rest of this paper is organized as follows. In Section II, a multi-interval economic dispatch problem is formulated and a time decomposition strategy is proposed to divide ED horizontally. In Section III, a distributed coordination strategy, which is based on auxiliary problem principle, and an initialization technique are presented to solve ED subproblems in a parallel manner. Simulation results illustrated in Section IV, and concluding remarks are provided in Section V.

## II. THE PROPOSED TIME DECOMPOSITION FRAMEWORK

Although the proposed algorithm can be applied to various multi-interval scheduling problems, for the sake of explanation, we focus on the multi-interval (i.e., weekly) economic dispatch problem in this paper.

### A. Economic Dispatching

The goal of the economic dispatching is to determine the power generation scheduling that leads to the lowest cost while supplying system and equipment constraints. The objective function is to minimize the summation of production costs of generating units. Equality constraints ($h$) include the nodal power balance (3), and inequality constraints ($g$) consist of limits of generating units (4), ramping up/down limits (5) and (6), limits of line flows (7), and system reserve requirements (8).

$$\min_{p_{u,t}} \sum_t \sum_u \underbrace{a_u \cdot p_{u,t}^2 + B_u \cdot p_{u,t} + C_u}_{f(p_{u,t})} \quad (1)$$

s.t.

$$P_L = SF(K_P \times P - K_D \times D) \quad (2)$$

$$K_L \times P_L = K_P \times P - K_D \times D \quad (3)$$

$$\underline{P_{u,t}} \le p_{u,t} \le \overline{P_{u,t}} \quad \forall u, \forall t \quad (4)$$

$$p_{u,t} - p_{u,t-1} \le UR_u \quad \forall u, \forall t \quad (5)$$

$$p_{u,t-1} - p_{u,t} \le DR_u \quad \forall u, \forall t \quad (6)$$

$$\underline{P_l} \le |PL| \le \overline{P_l} \quad \forall l \quad (7)$$

$$\sum_u p_{u,t\ max} \ge D_t + R_t \quad \forall t \quad (8)$$

where

$$h(p_{u,t}) = 0 \quad \rightarrow \{(2),(3)\}$$

$$g(p_{u,t}) \le 0 \quad \rightarrow \{(4),(5),(6),(7),(8)\}$$

$$u \in \{1, \dots, N_g\}, \qquad t \in \{1, \dots, N_t\}$$

Parameter $t$ denotes the time interval, and $u$ is the index for units. $K_L$, $K_D$ and $K_P$ are incident matrices of line, demand and generating units. SF denote the shift factor matrix. UR and DR are ramping up/down, D is demand vector. $R_t$ is the reserve requirement at interval $t$.

### B. Time decomposition

For solving steady-state analysis and decision making problems, the simplest way is to solve the problem in a centralized way for the whole operation horizon. The time horizon of such problems is shown in Figure 1 (a). However, as the duration of the horizon increases, the size of the problem and consequently the time of problem-solving and the computational burden increases nonlinearly.

In order to increase the speed of problem-solving, it is suggested to decompose the horizon into several sub-horizons with *n* equal intervals like hours, and then solve all subproblems in parallel. The sub-horizons of such problems is shown in Figure 1 (b). Equal intervals are more efficient since when we want to solve the problem in parallel, the size of the largest subproblem determines the time of problem-solving. Most methods that are being used, disregard the connections between time intervals because of simplicity.

Even some methods that consider temporal constraints of shared variables, usually add the constraints of the end of interval *i* to the beginning of interval *i*+1 as a hard constraint. In other words, the constraints of the end of interval *i* will be as initial values of the beginning of interval *i*+1. However, the initial value assigned for the interval *i*+1 might not be optimal for the optimization of interval *i*+1 and this can make the solution suboptimal.

However, if we do not consider the constraints of shared variables, the solution might be infeasible for application in real systems. In this study, the ramping is coupling the shared variables. Therefore, in order to have realistic solutions, we have to consider the ramping. With this aim, we propose to add an extra interval to the sub-horizons, which is called a coupling interval or complicating interval. Figure 1 (c) shows the coupling intervals, like *tci*.

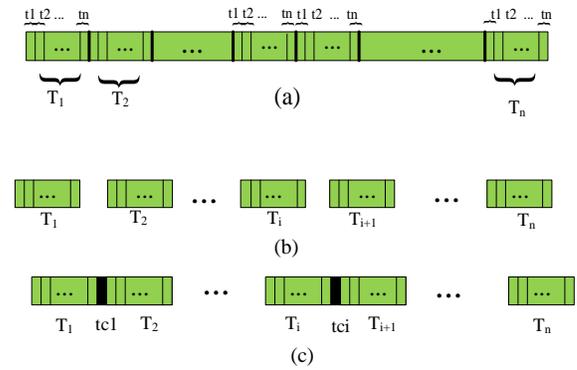

Fig. 1. (a) The problem is solved in a large time interval. (b) The time horizon of the problem is decomposed into several time intervals and each subproblem is solved separately. (c) The time horizon of the problem is decomposed into several time intervals and the coupling intervals are added to consider the connection between the intervals.

As an example, assume we want to solve an operation problem for two days. If we add the ramping constraint of hour 24 as the initial constraint of hour 1 of day 2 as a hard constraint, it is possible that in the view of day 2, it is not an optimal value of generating power and the solution is suboptimal. It is possible that day 2 wants to ask the hour 24 of day 1 to change its value of generating power to minimize the cost. These days should come to a compromise on the amount the generation of their units in the connecting hours considering ramp rate.

Therefore, we add an extra hour 25 to the last hour of day 1, which is optimized from the view of day 1 and it should be equal to the amount the power in hour 1 of day 2, which is optimized in the view of day 2. Days 1 and 2 should come to compromise about the amount of generated power in this hour as the optimization goes on.

### III. COORDINATION STRATEGY

The output of Section II is a set of optimization subproblems that are interconnected through the coupling intervals. If the coupling intervals are ignored, i.e., interdependencies of the sub-horizons are eliminated, the optimization subproblems can be solved independently. However, the coupling variables cannot be eliminated because of the ramp rate of the generating units. Thus, a strategy is needed to coordinate the subproblems and find the optimal and feasible solution form the perspective of the power systems and its components. Since our main goal is to reduce the computational burden and solution time, a parallel distributed algorithm is needed.

We model the coupling intervals with a set of complicating constraints and then convert these constraints to a set of coupling (also called shared or complicating) variables. A distributed coordination algorithm is presented based on the concept of auxiliary problem principle (APP) [15].

#### A. Auxiliary Problem Principle

APP is an iterative method aiming at finding the optimal solution of several coupled optimization problems in a distributed manner [15]. This method is based on the concept of augmented Lagrangian relaxation.

Assume that the considered scheduling interval is one week, and the problem is decomposed into seven sub-horizons each of which is one day. Consider two consecutive days $n$ and y $n+1$. Power output of the generating units at hour (i.e., time interval) 24 of day $n$ and hour 1 of day $n+1$ are linked together through the ramping rates of the units. We assume a coupling hour 25 for day $n$ that the amount of generated power (by each unit) in this hour must be the same as the amount of generated power in hour 1 of day $n+1$. Thus, generating powers at hour 25 are the shared variables between the two days. We denote the shared variables on day $n$ by $\phi_n$ and in day $n+1$ by $\phi_{n+1}$. Since $\phi_n$ and $\phi_{n+1}$ are physically the same, thus $\phi_n - \phi_{n+1} = 0$.

To achieve this consistency, we deploy APP that is an iterative approach. We formulate the following ED subproblem for day $n$ at iteration $k$:

$$\min_{(x_n^k, \Phi_n^k)} \sum_{u,t} f(p_{u,t}^k) \quad (9)$$
$$+ \left(\frac{p}{2}\left\|\Phi_n^k - \Phi_n^{*k-1}\right\|^2 + \gamma \Phi_n^{k\,\dagger}\left(\Phi_n^{*k-1} - \Phi_{n+1}^{*k-1}\right)\right.$$
$$\left. + \lambda^{(k-1)\,\dagger}\Phi_n^k\right)$$

s.t.

$$h_n(x_n^k, \Phi_n^k) = 0$$
$$g_n(x_n^k, \Phi_n^k) \leq 0$$
$$x_n^k = \{p_{u,t,n}^k\}, \Phi_n^k = \{p_{u,tc,n}^k\}, \Phi_{n+1}^{*k-1} = \{p_{u,tc,n+1}^{*k-1}\}$$

where $x_n$ is the set of power generated during day $n$, $\lambda^k$ is the vector of Lagrange multipliers at iteration $k$, and $\rho$ and $\gamma$ are suitable positive constants. $\Phi_n^{*k-1}$ and $\Phi_{n+1}^{*k-1}$ indicate the values of the shared variables of days $n$ and $n+1$ that are determined iteration $k$-1, an $\Phi_n^k$ is the shared variable of day $n$ that needs to be determined in iteration $k$. That is, $\Phi_n^{*k-1}$ and $\Phi_{n+1}^{*k-1}$ are known in (4) while $\Phi_n^k$ is a decision variable.

A similar ED subproblem is formulated for day $n+1$ as follows:

$$\min_{x_{n+1}^k, \Phi_{n+1}^k} \sum_{u,t} f(p_{u,t}^k) \quad (10)$$
$$+ \left(\frac{p}{2}\left\|\Phi_{n+1}^k - \Phi_{n+1}^{*k-1}\right\|^2 + \gamma \Phi_{n+1}^{k\,\dagger}\left(\Phi_{n+1}^{*k-1} - \Phi_n^{*k-1}\right)\right.$$
$$\left. - \lambda^{(k-1)\,\dagger}\Phi_{n+1}^k\right)$$

subject to the constraints of day $n + 1$. The penalty multiplier $\lambda$ needs to be updated at the end of each iteration according to

$$\lambda^k = \lambda^{k-1} + \alpha\left(\Phi_{n+1}^{*k} - \Phi_n^{*k}\right) \quad (11)$$

where $\alpha$ is a suitable positive constant. Note that the value of the Lagrange multiplier $\lambda$ in each iteration corresponds to the cost to maintain the consistency constraint. The above formulation can be generalized for a power system including multiple time intervals.

The pseudo code for the implementation to solve the APP-based distributed ED is given in Table I. Note that the considered ED problem is convex. APP is proven to converge to the global optimal solution under the convexity condition [15].

#### B. Initialization

In general, one of the main drawbacks of distributed/ decentralized optimization algorithms (such as APP) is their dependency on initial conditions. That is, the convergence performance of these algorithms may change if two different sets of initial conditions are used. If a set of good initial conditions is selected, APP will potentially converge in much fewer iterations compared to a case in which good initial conditions are not available. A good initial condition is system/problem dependent. This is an ongoing research in power systems and operations research communities.

TABLE I
The pseudo code for coordinating ED subproblems with APP

1: Decompose the considered horizon into *i* equal Sub-horizons
2: Initialize $\Phi_n^{*_0} \forall n, \lambda, \alpha, \rho$ and set $k=0$
3: **while** $\Phi_n^{*_k} - \Phi_{n+1}^{*_k} > \varepsilon, \ k = k+1$ **do**
4:   Solve the ED subproblems in parallel and determine the optimal values of $x_n$ and $\Phi_n^{*_k}$
5:   Exchange $\Phi_n^{*_k}$ between the subproblems
6:   Update $\lambda^k = \lambda^{k-1} + \alpha \left( \Phi_n^{*_k} - \Phi_{n+1}^{*_k} \right)$
7: **end while**

Since the main goal of the proposed time decomposition is to decrease the computational costs, we need to choose a suitable starting point. In this section, we take advantage of characteristics of the power system and propose a technique to find a set of good initialize conditions for the distributed ED algorithm. To initialize the problem, we ignore the coupling intervals (i.e., the sub-horizons are independent) and solve the optimization subproblems in parallel. Note that, in this paper, the considered horizon of ED is one week. Intuitively, since the load does not drastically change by the transition from the last time interval of sub-horizon $n$ to the first interval of sub-horizon $n+1$, ignoring ramping rates of the units (which are eliminated as the coupling intervals) does not impose a large error to the problem. Although the obtained results might not be feasible and optimal, they are merely close to the final results, which are optimal and feasible. We use results of this procedure to initialize the APP-based distributed economic dispatch problem.

The pseudo code for the distributed ED with the initialization technique is given in Table II.

TABLE II
The pseudo code for coordinating ED subproblems with APP + initialization

1: Decompose the considered horizon into *i* equal sub-horizons
2: Ignore the coupling time intervals
3: Solve the ED subproblems in parallel
4: Use the obtained results to initialize the shared variables $\Phi_{n+1}^{*_0}, \Phi_n^{*_0}$
5: Set multipliers $\lambda, \alpha, \rho$ and set $k=0$
6: **while** $\Phi_{n+1}^{*_k} - \Phi_n^{*_k} > \varepsilon, \ k = k+1$ **do**
7:   Solve the ED subproblems in parallel and determine the optimal values of $x_n$ and $\Phi_n^{*_k}$
8:   Exchange $\Phi_n^{*_k}$ and $\Phi_{n+1}^{*_k}$ between the subproblems
9:   Update $\lambda^k = \lambda^{k-1} + \alpha \left( \Phi_n^{*_k} - \Phi_{n+1}^{*_k} \right)$
10: **end while**

## IV. CASE STUDY AND NUMERICAL RESULTS

The proposed algorithm is applied to solve a week-ahead ED problem on the IEEE 118-bus test system. The operation horizon is divided into seven sub-horizons, each including 24 intervals (i.e., each sub-horizon is one day).

All simulations are carried out using YALMIP toolbox in Matlab [16] and ILOG CPLEX 12.4's QP solver on a 3.7 GHz personal computer with 16GB RAM.

Three cases are studied:

- Case 1: Centralized system scheduling
- Case 2: The proposed distributed algorithm without the initialization technique
- Case 3: The proposed distributed algorithm with the initialization technique

*Case 1:* The conventional centralized algorithm is deployed to solve the problem. The obtained results are used as the benchmark to validate results of the proposed distributed algorithm. Table III shows the computational time and operation costs. The centralized ED converges after 1.8 seconds. The operation cost is 12.343 million dollars.

*Case 2*: The proposed distributed algorithm is used to solve the ED problem. In this case, the initial values of the shared variables are set to zero. As shown in Table III and Fig. 2, the algorithm converges after 11 iterations within 1.7 seconds, which is slightly less than that of the centralized ED. However, the operation cost is 0.3236% larger than the centralized ED. As an example, Fig. 3 shows the power generated by unit 42, which is a shared variable between days 1 and 2, over the course of iterations. Note that in several intervals, these variables are the same; however, the algorithm stops when all shared variables are the same.

*Case 3*: The proposed distributed algorithm is used with the suggested initialization technique. The algorithm converges after 3 iterations (2 APP iterations plus the initiation step) within 0.4 seconds, which is almost 70% faster than the centralized ED. In addition, the relative error of the operation costs obtained by the decentralized and centralized approaches is almost zero. In Fig. 4, we show an example of the shared variables between days 1 and 2. These shared variables are power produced by unit 10 in hour 1 of day 2. The difference between the shared variables is zero at iteration 2. Note that if we consider the initialization step, the algorithm converges after 3 iterations.

TABLE III
RESULTS OF THE THREE CASES

| Case No | Total cost ($) | Iteration | Relative Error % | Overall time (s) |
|---|---|---|---|---|
| **Case 1** | $1.2343 \times 10^7$ | - | - | 1.85 |
| **Case 2** | $1.2383 \times 10^7$ | 11 | 0.3236 | 1.7 |
| **Case 3** | $1.2343 \times 10^7$ | 3 | $5.9562 \times 10^{-5}$ | 0.4 |

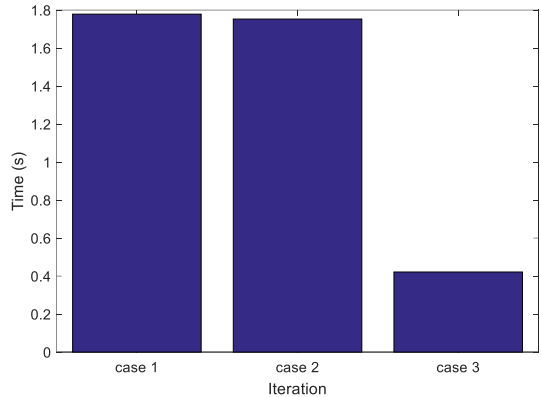

Fig. 2. Comparing overall time of centralized algorithm and proposed distributed algorithms.

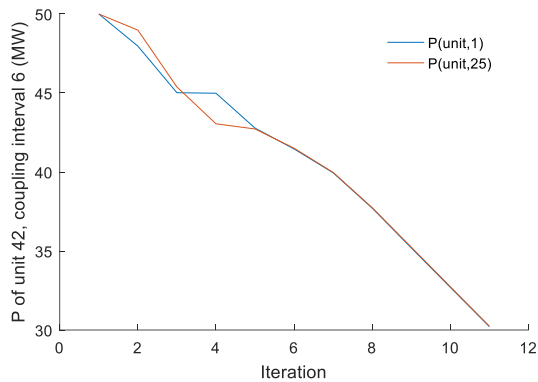

Fig. 3. Shared variable corresponding to unit 42 (between days 1 and 2) over the course of iterations.

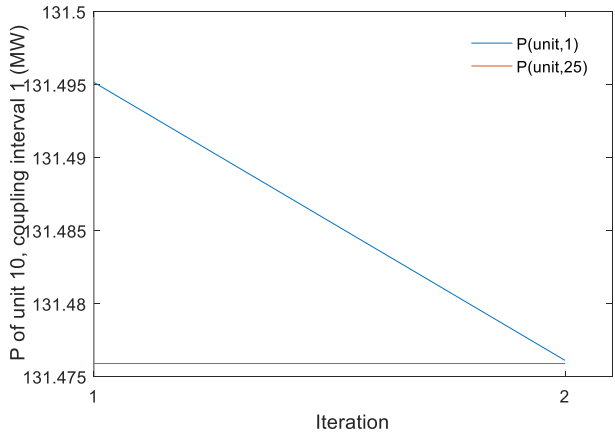

Fig. 4. Shared variable corresponding to unit 10 (between days 1 and 2) over the course of iterations.

## V. CONCLUSION

In this paper, we proposed a method to decompose the ED problem over the time horizon. An ED subproblem was formulated for each sub-horizon taking into account the ramp rates of the generating units. To mathematically model this point, a coupling interval was defined between two consecutive sub-horizons. The auxiliary problem principle approach was adopted to solve the ED subproblems in a parallel manner. To enhance the convergence speed, an initiation technique was presented. The results of the IEEE 118-bus test system showed the effectiveness of the proposed time decomposition-coordination framework to reduce the computational cost of the multi-interval ED problem by around 70%.